\documentclass{ws-ijmpd}
\usepackage[super,compress]{cite}
\usepackage{amsmath,amssymb}

\DeclareMathOperator\Tr{tr}
\newcommand\hash{\mathop{\#}}

\begin{document}

\markboth{C.~Beetle, J.S.~Engle, M.E.~Hogan, and P.~Mendon\c{c}a}
{Diffeomorphism invariant cosmological symmetry in full quantum gravity}

%
%

\title{Diffeomorphism invariant cosmological symmetry in full quantum gravity}

\author{Christopher Beetle${}^1$\footnote{cbeetle@fau.edu}, 
	Jonathan S. Engle${}^1$\footnote{jonathan.engle@fau.edu}, 
	Matthew E.~Hogan${}^{1,2}$\footnote{mhogan@wells.edu}, 
	Phillip Mendon\c{c}a${}^1$\footnote{pmendon1@fau.edu}}

\address{${}^1$ Department of Physics, Florida Atlantic University, 777 Glades Road,\\
Boca Raton, Florida 33431, USA}

\address{${}^2$ Department of Mathematical and Physical Sciences, Wells College, 170 Main Street,\\
Aurora, New York 13026, USA}


%

\maketitle


\begin{abstract}
This paper summarizes a new proposal to define rigorously a sector of loop quantum gravity at the diffeomorphism invariant level corresponding to homogeneous and isotropic cosmologies, thereby enabling a detailed comparison of results in loop quantum gravity and loop quantum cosmology.  The key technical steps we have completed are (a) to formulate conditions for homogeneity and isotropy in a diffeomorphism covariant way on the classical phase space of general relativity, and (b) to translate these conditions consistently using well-understood techniques to loop quantum gravity.  Some additional steps, such as constructing a specific embedding of the Hilbert space of loop quantum cosmology into a space of (distributional) states in the full theory, remain incomplete.  However, we also describe, as a proof of concept, a complete analysis of an analogous embedding of homogeneous and isotropic loop quantum cosmology into the quantum Bianchi I model of Ashtekar and Wilson-Ewing.  Details will appear in a pair of forthcoming papers.
\end{abstract}




\section*{}

Symmetry is a powerful tool for simplifying the physics of a system in order to extract predictions.  In the case of quantum gravity, the most promising place to seek potentially observable predictions is in cosmology, where classical gravity breaks down at the Big Bang and quantum effects become important.  The universe appears spatially homogeneous and isotropic to a very good approximation at large scales, and cosmological models exploit this symmetry to enable exact solutions of the otherwise prohibitively complicated dynamics of Einstein's theory. 

Imposing homogeneity and isotropy, or indeed any spatial symmetry, in \textit{quantum} gravity, however,  immediately runs into two obstacles: 
\begin{enumerate}
\item Symmetry must be imposed on the full set of initial data --- both configuration and momentum variables.  Classically this is necessary for the future evolution to continue to be symmetric.  However, at the quantum level, this violates the Heisenberg uncertainty principle.
\item Usually symmetry is imposed by requiring invariance of the initial data under a fixed action of the symmetry group.  However, such a method of imposing symmetry necessarily breaks the diffeomorphism invariance of general relativity.  As a consequence, in the quantum theory, such a method of imposing symmetry will never be consistent with imposition of the dynamical constraints.\footnote{Unless one quantizes the full theory by fixing the diffeomorphism gauge first.  However, such a strategy is not usually proposed as a way to obtain a fundamental theory of quantum gravity, due to known problems which occur when gauge fixing is used to obtain a non-perturbative quantization of a gauge theory with non-Abelian constraints \cite{gribov1978}.}
\end{enumerate}

In this paper, we introduce a way to impose symmetry in quantum gravity that overcomes these fundamental obstacles.  Specifically, we formulate new \textit{diffeomorphism covariant} constraints that impose homogeneity and isotropy in the phase space of general relativity.  These constraints involve complex functions on phase space, which have a \textit{closed} Poisson algebra.  The latter fact allows us to overcome the first obstacle mentioned above in a manner similar to that used in the Gupta--Bleuler quantization of the electromagnetic field\cite{gupta1950, bleuler1950}.  Furthermore, the complex constraint functions can be quantized using existing techniques in the framework of loop quantum gravity.  This yields for the first time a systematically motivated homogeneous and isotropic sector at the diffeomorphism-invariant level of the quantum theory.

Our ultimate goal is to use this homogeneous and isotropic sector of loop quantum gravity to develop a clear and detailed connection to homogeneous and isotropic loop quantum cosmology.  The latter arises by imposing the symmetry classically, prior to quantization.  The resulting quantum model is much more tractable than the full theory, and allows explicit comparison between theoretical predictions and observations, but loses any \textit{a priori} connection to the full quantum theory\relax
\footnote{See, however, recent progress in Refs.~\refcite{bodendorfer2015,ac2015,ac2014,fleischhack2010,hanusch2014,hanusch2013,engle2013,engle2007}{}.}{}\relax
.  Our aim is to restore that connection, so that observational constraints on a quantum cosmological model can also help dictate choices that must be made in quantizing the full theory.  In particular, one would like to be able to use constraints on the dynamics of the quantum cosmological model to similarly constrain the dynamics of the full theory.

Some steps of our program to relate a symmetric sector of loop quantum gravity to loop quantum cosmology remain incomplete, however.  We will summarize the steps that have been completed below, and mention those that are the subject of ongoing work.  We have, however, completed the analogous task relating the loop quantum Bianchi I model of Ashtekar and Wilson-Ewing \cite{aw2009} to homogeneous and isotropic loop quantum cosmology.  We were surprised to find that the homogeneous and isotropic model could be embedded in the Bianchi I model using a known linear mapping.  Specifically, the embedding is the adjoint of the projection from Bianchi I to the fully symmetric model previously found by Ashtekar and Wilson-Ewing.  We will discuss this simplified embedding scheme, both as a proof of concept for our larger goal and as an interesting result in its own right.
 
Details of our results will be published in two forthcoming papers \cite{behm2016, behm2016a}.

\section{Definition of the symmetry constraints}

The basic variables of loop gravity are an $SU(2)$ connection $A^i_a$ and a densitized triad $\tilde{E}^a_i$.  These are related to the traditional ADM phase-space variables --- the spatial metric and extrinsic curvature $q_{ab}, K_{ab}$ --- by 
\begin{align*}
A^i_a = \Gamma^i_a + \gamma K^i_a \qquad \text{and} \qquad
\tilde{E}^a_i = | {\det e} |\,  e^a_i
\end{align*}
where $q_{ab} = e^i_a e_{bi}$, $e^a_i$ is the inverse of $e^i_a$, $\Gamma^i_a$ is the spin connection determined by the triad $e^i_a$, $K^i_a := K_{ab} e^{ai}$, and $\gamma$ is the Barbero-Immirzi parameter\cite{barbero1995, immirzi1997}, 
which usually is fixed through considerations of black hole entropy\cite{dl2004, meissner2004}.

We seek a condition on these variables equivalent to homogeneity and isotropy that is nevertheless \textit{diffeomorphism invariant}.  An initial data set $(q_{ab}, K_{ab})$ is homogeneous and isotropic if and only if  $q_{ab}$ is \textit{maximally symmetric} --- that is, possesses six Killing vector fields --- and $K_{ab}$ is proportional to $q_{ab}$ by a constant: 
\begin{align}
\label{Kabsymm}
K_{ab} = H q_{ab}.  
\end{align}
It is well known that if $q_{ab}$ is maximally symmetric, then its Riemann curvature has the form 
\begin{align}
\label{maxsymmR}
R_{abcd} = \frac{1}{6}\, R\, \bigl( q_{ac}\, q_{bd} - q_{ad}\, q_{bc} \bigr), 
\end{align}
where the scalar curvature $R$ is constant over space.  In fact, the converse is also true \cite{behm2016}.  It follows that a spacetime admits a foliation by maximally symmetric spatial slices if and only if it admits initial data satisfying 
\begin{align}
\label{triadsymm}
F_{ab}{}^i(\Gamma) = \rho\, \Sigma_{ab}{}^i \qquad \text{and} \qquad K^i_a = H\, e^i_a 
\end{align}
for some constants $\rho$ and $H$, where $F_{ab}{}^i(\Gamma)$ is the curvature of $\Gamma^i_a$
and $\Sigma_{ab}^i := \epsilon^i{}_{jk}e^j_a e^k_b$.

The two conditions (\ref{triadsymm}) can be combined in a useful form for quantization using the  complexifier method of Thiemann \cite{thiemann2002}.  Fixing a complexifier function $C$ on phase space, one defines a 1-parameter family 
\begin{equation}
	\frac{\partial\,}{\partial t} {}^t \mathbf{O} := \mathrm{i}\, \bigl\{ {}^t \mathbf{O}, C \bigr\}
	\qquad\text{with}\qquad
	{}^0 \mathbf{O} := O
\end{equation}
of complex functions ${}^t \mathbf{O}$ on phase space deriving from a given (generally real-valued) function $O$.  Importantly, the corresponding family of operators in the quantum theory will have the form 
${}^t \hat{\mathbf{O}} = \mathrm{e}^{-t \hat C}\, \hat O\, \mathrm{e}^{ t \hat C}$, 
so in principle it is straightforward to quantize such complexified observables.

Choosing $C$ to be proportional to the total volume $V$ of space and setting $O = A_a^i$ yields a complexified connection 
\begin{equation}
	{}^t\! \mathbf{A}_a^i = A_a^i + \mathrm{i} \alpha t\, e_a^i, 
\end{equation}
where $\alpha$ is a constant with units of inverse length.  The (co-)triad $e_a^i$ is invariant under this complexification flow.  Calculating the curvature of this complexified connection shows that the real and imaginary parts of the single relation 
\begin{equation}
	\mathbf{F}_{ab}{}^i = \mathbf{b}\, \Sigma_{ab}{}^i, 
\end{equation}
where we have set $t = 1$ and $\mathbf{b}$ is a complex constant, imply \textit{both} of the real symmetry conditions (\ref{triadsymm}).  Furthermore, taking a wedge product with $f_{ij}\, e_c^j$, where $f_{ij}$ is an arbitrary smearing function, and integrating over all space gives 
\begin{equation}\label{BVdef}
	\mathbf{B}[f] := \int f_{ij}\, \mathbf{F}^i \wedge e^j
		= \mathbf{b} \int f_{ij}\, \delta^{ij}\, \epsilon_{klm}\, e^k\, e^l\, e^m =: \mathbf{b}\, V[f].
\end{equation}
The right side here is proportional to the proper-volume integral of the trace of $f_{ij}$ over all space.  Cross-multiplying two such equations allows us to eliminate the constant $\mathbf{b}$ and write the symmetry conditions in the form 
\begin{equation}\label{Sdef}
	\mathbf{S}[f, g] := \mathbf{B}[f]\, V[g] - \mathbf{B}[g]\, V[f] = 0 
\end{equation}
for all $f_{ij}$ and $g_{kl}$.

The non-complexified version of $B[f]$ is nearly identical to the Euclidean self-dual Hamiltonian constraint of general relativity, which arises when $f_{ij} = N\, \delta_{ij}$.  It can be quantized using the same basic method used for the Euclidean Hamiltonian constraint.  Finally, we can complexify the result in the quantum theory by setting 
\begin{equation}
\label{complexifyB}
	\hat{\mathbf{B}}[f] := \mathrm{e}^{-\hat V / v_0}\, \hat B[f]\, \mathrm{e}^{\hat V / v_0}, 
\end{equation}
where $v_0$ is an fixed, but arbitrary, constant with units of volume.

%
%

The functions in equation
(\ref{Sdef}) are our constraints, whose vanishing is equivalent to homogeneity and isotropy of the gravitational field
with respect to \textit{some} action of a maximal symmetry group. 
They are manifestly diffeomorphism and gauge covariant, so that their vanishing for \textit{all} test functions
is a diffeomorphism and gauge \textit{invariant} condition. 



\section{First class nature}

A necessary condition for a given set $\bigl\{ C_i \bigr\}$ of classical constraints to admit operator analogues $\bigl\{ \hat C_i \bigr\}$ that can be imposed strongly (\textit{i.e.}, $\hat C_i\, | \psi \rangle = 0$) and consistently (\textit{i.e.}, simultaneously) at the quantum level is that their classical Poisson algebra should close.  That is, there should exist phase-space functions $\lambda_{ij}{}^k$ such that 
\begin{equation}
	\bigl\{ C_i, C_j \bigr\} = \sum_k \lambda_{ij}{}^k\, C_k.
\end{equation}
We refer to such constraints as a \textit{first-class set}.

We wish to impose the symmetry conditions $\mathbf{S}[f,g]$ discussed above strongly in quantum theory, as well as the Gauss and diffeomorphism constraints $C[\Lambda]$ and $C[\vec{N}]$ of general relativity.  Remarkably, these functions \textit{do} form a first class set classically.  The first thing to check is that the symmetry conditions form a first class set by themselves.  One finds 
\begin{align}
	\bigl\{ \mathbf{S}[f, g], \mathbf{S}[h, k] \bigr\} 
		&= \frac{\kappa\gamma}{2}\, V[g]\, \mathbf{S}[f \hash h, k] 
			+ \frac{\kappa\gamma}{2}\, \mathbf{S}[g, k] \int \Tr f\, e^k 
				\wedge \mathbf{D} \bigl( h_{kl}\, e^l \bigr)
			\\&\hspace{10em}\notag
			- (f \leftrightarrow g) 
			- (h \leftrightarrow k)
			+ (fh \leftrightarrow gk), 
\end{align}
where $\mathbf{D}_a$  denotes the gauge-covariant derivative defined by $\mathbf{A}_a^i$, 
and where we have defined the ``hash product'' 
\begin{equation}
	(f \hash h)_{ij} := 2\, f_{ik}\, V^{abc}\, e^k_a\, \mathbf{D}^{}_b \bigl( h_{jl}\, e^l_c \bigr) 
		- f_{ij}\, V^{abc}\, e_a^k\, \mathbf{D}^{}_b \bigl( h_{kl}\, e^l_c \bigr) 
		- (f \leftrightarrow h)
\end{equation}
of smearing functions.  The next thing to check is the algebra of the symmetry conditions and the constraints.  One finds immediately that 
\begin{equation}
	\bigl\{ C[\Lambda], \mathbf{S}[f, g] \bigr\} = 0
	\qquad\text{and}\qquad
	\bigl\{ C[\vec N], \mathbf{S}[f, g] \bigr\} 
		= \mathbf{S}[\mathcal{L}_{\vec N} f, g] + \mathbf{S}[f, \mathcal{L}_{\vec N} g]
\end{equation}
because $\mathbf{S}[f, g]$ is (Gauss) gauge-invariant and diffeomorphism-covariant.  The total algebra of symmetry conditions and constraints therefore closes.  Thus, it is reasonable to seek a sector of (distributional) states in loop quantum gravity lying simultaneously in the kernels of all the symmetry condition operators, as well as all the Gauss and diffeomorphism constraint operators.

\section{Quantum implementation}

To construct an operator corresponding to the smeared constraint function $\mathbf{S}[f,g]$ (\ref{Sdef}), one first constructs operators corresponding to $\mathbf{B}[f]$ and $V[g]$ in (\ref{BVdef}).  This is a simple task in the framework of loop quantum gravity.  First, the standard quantization of the volume element given in Ref.~\refcite{al1997} yields a canonical definition for $\hat V[f]$.  Second, as we mentioned above, $\mathbf{B}[f]$ has a form closely related to that of the Euclidean self-dual Hamiltonian constraint 
\begin{align*}
C_{\rm Eucl}[N] =  \int N F^i \wedge e_i, 
\end{align*}
where $F^i_{ab}$ is the curvature of $A^i_a$.  Specifically, $\mathbf{B}[f]$ differs only in that (a) the two indices are no longer contracted with each other, but rather with those of a matrix-valued smearing function, and (b) one uses the complexified connection $\mathbf{A}^i_a$ instead of the real connection $A^i_a$.  Neither of these differences affect Thiemann's procedure \cite{thiemann1996} for quantizing $C_{\rm Eucl}[N]$.  Exactly the same methods can be used to define an operator $\hat{\mathbf{B}}[f]$.
One can either carry out the Thiemann strategy using the quantum complexified connection directly, 
or first using the quantum real connection and then complexifying as in (\ref{complexifyB}). The resulting operator in either case 
will be the same due to the structure of quantum complexification.

There is a further subtlety to be addressed, even with $\hat{\mathbf{B}}[f]$ and $\hat{V}[f]$ constructed.  
As in the case of Thiemann's Hamiltonian constraint, the operator $\hat{\mathbf{B}}[f]$ has a well-defined action only within the space of solutions $\mathcal{V}_{\rm diff}$ to the quantum diffeomorphism constraint.  Furthermore, as with Thiemann's constraint (when it is smeared with a non-constant lapse), $\hat{\mathbf{B}}[f]$ maps  diffeomorphism-invariant states  out of $\mathcal{V}_{\rm diff}$.  That is, its domain and range are mutually exclusive.  In contrast, the smeared volume operator $\hat{V}[f]$ is well-defined on \textit{all} states, diffeomorphism invariant or not, 
but generally maps diffeomorphism-invariant states out of $\mathcal{V}_{\rm diff}$.  
These facts determine the operator ordering to use when building the operators $\hat{\mathbf{S}}[f, g]$
analogous to (\ref{Sdef}).  Namely, if we choose the domain of $\hat{\mathbf{S}}[f, g]$ to be $\mathcal{V}_{\rm diff}$, then we must order $\hat{\mathbf{B}}[f]$ to act before $\hat{V}[g]$, and likewise $\hat{\mathbf{B}}[g]$ to act before $\hat{V}[f]$.  The space of states annihilated by $\hat{\mathbf{S}}[f,g]$ for all smearing functions $f^i{}_j$ and $g^i{}_j$ then defines our \textit{homogeneous and isotropic sector} $\mathcal{V}_{\rm symm} \subset \mathcal{V}_{\rm diff}$.

\section{Program to relate dynamics}
\label{dynamics}

The Gauss and diffeomorphism constraints hold identically in loop quantum cosmology, and also in the new homegeneous and isotropic sector of loop quantum gravity described above. The only remaining constraint is the (Lorentzian) Hamiltonian constraint $C[N]$.  Furthermore, there is only one independent smearing of the Hamiltonian constraint that needs to be imposed in the homogeneous and isotropic context, which can be taken to be the unit-lapse smearing $C[1]$.  There exists a quantum analogue of $C[1]$ in both loop quantum cosmology and loop quantum gravity, which we denote $\hat{C}_S$ and $\hat{C}[1]$, respectively.  One of the key aims of our program is to give a direct relation between these two operators at the quantum level.  

This part of our program remains incomplete, but the general strategy we intend to follow is to construct an embedding $\iota$ of LQC states into $\mathcal{V}_{\rm symm}$ such that, at least in an approximate sense, the matrix elements of the two Hamiltonian constraint operators match:
\begin{align}
\label{matel}
\frac{\langle \iota \psi, \hat{C}[1] \iota \phi \rangle}
{\sqrt{\langle \iota \psi, \iota \psi \rangle \langle \iota \phi, \iota \phi \rangle}} 
= 
\frac{\langle \psi, \hat{C}_S \phi \rangle}
{\sqrt{\langle \psi, \psi \rangle \langle \phi, \phi \rangle}}.
\end{align}
This condition should hold for all LQC states $\psi$ and $\phi$. 
The division by the norms here is needed because $\iota$ will in general not be norm-preserving, 
so that $\phi$, $\psi$, $\iota \phi$ and $\iota \psi$ will in general not be simultaneously normalizable.
The extent to which the above condition can be satisfied for some $\iota$ is then a measure of the extent
to which the two definitions of dynamics are consistent.
Alternatively, $\iota$ can be determined by imposing a similar equality of matrix elements for one or more 
pairs of simpler operators.  Then, given an operator $\hat{C}[1]$ in the full theory, equation (\ref{matel})
will uniquely determine a corresponding operator in LQC, which can then be compared with the
Hamiltonian constraint operator used in the LQC literature.

In practice, the above prescriptions are more involved than they sound, 
as $\mathcal{V}_{\rm symm}$  is expected to consist in states which are not normalizable
in the diffeomorphism invariant inner product on $\mathcal{V}_{\rm diff}$\cite{almmt1995, al2004}, 
so that the left hand side of the above equation will likely
be of the form $\infty/\infty$ and so will need to be regularized.  
We refer to the larger paper \cite{behm2016} for such details.

\section{Application to Bianchi I}

Even though the program of section \ref{dynamics} has not been carried to completion in relating the full theory of LQG to LQC, it \textit{has} been completed\cite{behm2016a} to impose isotropy at the quantum level in a Bianchi I model \cite{aw2009}.  This is a simpler, but still non-trivial, problem.  This application allows a detailed comparison of the isotropy-reduced quantum dynamics of Bianchi I \cite{aw2009} with that of the well-established isotropic LQC model \cite{aps2006, abl2003}.
 The results were unexpected.  We found an exact match of kinematics and dynamics in the two theories,
 as well as an exact match of naturally defined operators in each model corresponding to spatial volume, 
 average spatial curvature, and Hubble rate. 
 Furthermore, the embedding $\iota$ in this application
 to Bianchi I turned out to be none other than the adjoint of the dynamical
 projector $\hat{\mathbb{P}}$ from Bianchi I to isotropic LQC proposed 
by Ashtekar and Wilson-Ewing in Ref.~\refcite{aw2009}.
That is, the image of the adjoint of their dynamical projector satisfies precisely
the symmetry constraint operator equation proposed in this paper, specialized to Bianchi I in a straightforward way.
If one includes scalar matter in the analysis, all of the above statements are again true, as long as the 
matter Hamiltonian depends on the geometry only via the volume, as is the case for the matter considered
in Ref.~\refcite{aw2009}.

In light of the successful application to Bianchi I, and of the equivalence to the t is dynamical projector method of 
Ref.~\refcite{aw2009} 
in that case, one can view the present program as way to reformulate the dynamical projector method and 
generalize it to the full theory of LQG.  What makes this possible is the sharpened understanding of the role of homogeneity and isotropy.

A further benefit of the application to Bianchi I is that, in matching operators in the two models, 
such as those corresponding to spatial 
curvature and Hubble rate, the ambiguity in how the operators are quantized is necessarily restricted. 
If one similarly tries to match operators corresponding to, for example, components of the four dimensional Ricci tensor,
one again expects a preferred class of quantizations to be selected.
It is worth checking whether the use of such quantizations of Ricci might resolve the issue
of apparent non-zero Ricci components in vacuum \mbox{Bianchi I} discussed in Ref.~\refcite{singh2016}.

\section*{Acknowledgements}
The authors are grateful to Ted Jacobson, Atousa Chaharsough Shirazi, Brajesh Gupt, Jorge Pullin, Parampreet Singh, Xuping Wang, and Shawn Wilder for discussions.
This work was funded in part by the U.S. National Science Foundation under grants PHY-1205968 and PHY-1505490,
and by NASA through the University of Central Florida's NASA-Florida Space Grant Consortium.

%

\end{document}